\documentclass[pdflatex,sn-mathphys]{sn-jnl}
\usepackage{soul}

\jyear{2021}%
\theoremstyle{thmstyleone}%
\theoremstyle{thmstyletwo}%
\theoremstyle{thmstylethree}%
\begin{document}

\title[ ]{Shaping Exciton Polarization Dynamics in 2D Semiconductors by Tailored Ultrafast Pulses}

\author[1,2]{\fnm{Omri} \sur{Meron}}\email{omrimeron@tauex.tau.ac.il}

\author[1,2]{\fnm{Uri} \sur{Arieli}}

\author[1,2]{\fnm{Eyal} \sur{Bahar}}

\author[1]{\fnm{Swarup} \sur{Deb}}

\author[1]{\fnm{Moshe} \sur{Ben-Shalom}}

\author[1,2]{\fnm{Haim} \sur{Suchowski}}

\affil[1]{\orgdiv{Condensed Matter Physics Department, School of Physics and Astronomy, Faculty of Exact
Sciences}, \orgname{Tel Aviv University}, \orgaddress{\city{Tel‐Aviv}, \postcode{6997801}, \state{Israel}}}

\affil[2]{\orgdiv{Center for Light‐Matter Interaction}, \orgname{Tel Aviv University}, \orgaddress{\city{Tel‐Aviv}, \postcode{6997801}, \state{Israel}}}


\abstract{
\textcolor{black}{
The ultrafast formation of strongly bound excitons in two-dimensional semiconductors provide a rich platform for studying fundamental physics as well as developing novel optoelectronic technologies. While extensive research has explored the excitonic coherence, many-body interactions, and nonlinear optical properties, the potential to study these phenomena by directly controlling their coherent polarization dynamics has not been fully realized.
In this work, we use a sub-10fs pulse shaper to study how temporal control of coherent exciton polarization affects the generation of four-wave mixing in monolayer $\text{WSe}_\text{2}$ under ambient conditions. By tailoring multiphoton pathway interference, we tune the nonlinear response from destructive to constructive interference, resulting in a 2.6-fold enhancement over the four-wave mixing generated by a transform-limited pulse. This demonstrates a general method for nonlinear enhancement by shaping the pulse to counteract the temporal dispersion experienced during resonant light-matter interactions. Our method allows us to excite both 1s and 2s states, showcasing a selective control over the resonant state that produces nonlinearity. By comparing our results with theory, we find that exciton-exciton interactions dominate the nonlinear response, rather than Pauli blocking. This capability to manipulate exciton polarization dynamics in atomically thin crystals lays the groundwork for exploring a wide range of resonant phenomena in condensed matter systems and opens up new possibilities for precise optical control in advanced optoelectronic devices.}

}

\maketitle

\section{Introduction}\label{sec1}
Crystalline semiconductors are integral components of today's electronic devices and are thus of fundamental significance to science and technology alike. 
A detailed understanding of their optical response, stemming from the complex collective dynamics of their periodically arranged charge carriers, is crucial for exploiting their interactions.
Optical coherent control through pulse-shaping is a well-developed method for approaching this challenge, aiming to optimize the \textcolor{black}{multiphoton pathway} interference between charge carrier excitations by pre-designed broadband pulses \cite{Goswami2003}. Indeed, optical coherent control applied to atoms, molecules \cite{Silberberg2009}, and quantum dots \cite{Brash2017}, characterized by well-defined energy levels and coherence times exceeding the picosecond timescale, has resulted in \textcolor{black}{enhancement of the transient population of an excited level \cite{Dudovich2002a}, selective quantum state control \cite{Dudovich2002}, creation and dissociation of chemical bonds \cite{Chelkowski1990, Assion1998, Levin2015}, etc.}
In the case of crystalline 3D semiconductors, however, the numerous excited states are separated by minute energy levels, which makes room-temperature coherent control extremely challenging.

Layered 2D semiconductors are positioned between these two extremes. The ultimate out-of-plane confinement and the planner crystalline periodicity make a perfect playground to explore electronic orders and unique optical responses. Mono-layer transition metal dichalcogenides (TMDs), in particular, exhibit a direct energy gap that hosts remarkably stable coulomb-bound electron-hole pairs referred to as excitons.   
These quasiparticles, which dominate the crystals' optical properties, even at room temperature, exhibit a Rydberg series of resonant excitonic transitions with decreasing oscillator strengths \cite{Chernikov2014, Mueller2018, Brem2019}. The first and particularly prominent $A1s$ orbital state of the band-edge exciton lays in the visible to near-infrared optical range and is identified by its relatively high quantum yield \cite{Mak2010, Amani2015}. This combination of unique semiconductor physical properties and heterostructure stacking capabilities \cite{Geim2013, Ciarrocchi2022}, point TMDs as promising, versatile, and compact building blocks for future optoelectronic devices \cite{Mueller2018, Chaves2020}.

\textcolor{black}{While various experimental methods such as pump-probe and multi-dimensional spectroscopy have effectively explored the coherent and nonlinear aspects of exciton \textcolor{black}{polarization} 
dynamics in TMDs \cite{Moody2015, Poellmann2015, Dey2016, Jakubczyk2016, Steinleitner2017, Trovatello2020, Li2021a, Bauer2022, Purz2022, Luo2023, Rodek2023}, the application of pulse shaping in this context remains largely untapped. 
} 
\textcolor{black}{
Unlike conventional techniques, which rely on scanning time delays between discrete pulses and are constrained by the pulses' temporal width in their transform-limited form, pulse shaping enables a smooth spectral phase modulation of a single pulse, adding a new level of control to these experiments. This approach creates a finely tailored color sequence that not only probes but also precisely controls the nonlinear coherent polarization dynamics. 
} 
\textcolor{black}{In this study, we introduce a novel method for selectively controlling and isolating nonlinear contributions from transient exciton populations within their coherent motion. 
This approach not only sheds new light on the contributions of many-body effects to the parametric nonlinear wave-mixing response but also demonstrates the potential of pulse shaping for precise quantum state control and tailored manipulation of nonlinear optical processes in layered 2D semiconductors.
}

Our results reveal that, under ambient conditions, the A1s exciton resonance plays a significant role in the observed temporal dynamics, driving the four-wave-mixing (FWM) response. \textcolor{black}{We establish a novel nonlinear enhancement method through constructive multiphoton pathway interference \cite{Dudovich2001, Bahar2022}, which is broadly applicable to resonant systems and complements several existing enhanced nonlinear effects in TMDs \cite{Dai2021}. Specifically, for the FWM generated by the A-exciton, our apparatus, which involves only phase shaping, increases the FWM intensity by \textcolor{black}{a factor of 2.6} 
compared to a transform-limited (TL) pulse.} Additionally, we demonstrate that by employing destructive interference, we can nearly eliminate the FWM signal while maintaining \textcolor{black}{high peak intensity} sub-10 fs pulses. The application of the TMD Bloch equations of motion successfully captures our experimental observations, highlighting the significant contribution of exciton-exciton (X-X) interactions to FWM signal generation. Furthermore, our observations indicate that the A1s exciton does not play a dominant role in sum frequency generation (SFG). Finally, we showcase the ability to control the contribution of the sequential A2s resonant state to the overall FWM signal, enabling simultaneous control over both A1s and A2s excited states. This method introduces a unique and selective control mechanism for coherent exciton responses in TMDs and their heterostructures, representing a significant advancement with potential applications in on-chip ultrafast and nonlinear optical technologies, including enhanced nonlinear sensing \cite{Malard2021}, compressed broadband light sources \cite{Liu2017}, and efficient frequency converters \cite{Trovatello2021}.

\section{Results}\label{sec2}

\begin{figure}[h]
    \centering
    \includegraphics[scale=0.47]{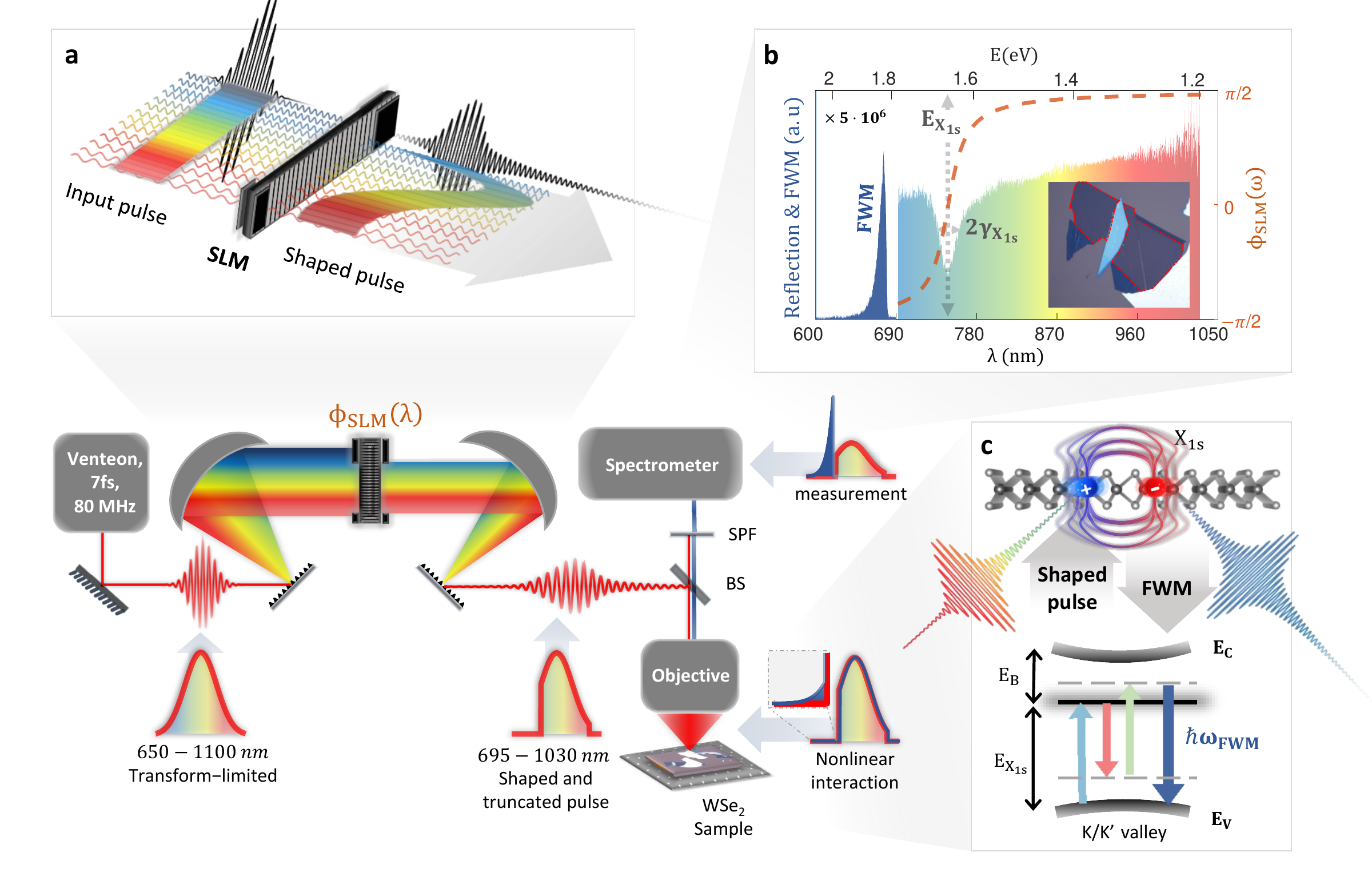}
    \caption{\textbf{Coherent control of the A-exciton resonance in monolayer $\text{WSe}_\text{2}$.} Experimental setup, an ultra-broadband sub-$10fs$ pulse ($650-1100nm$) propagates through an SLM situated in the Fourier plane of a 4f system, also used as a spectral edge filter. The shaped and truncated pulse (690-1100nm) is then focused on a sample, generating a nonlinear signal which is collected via reflection and redirected into a spectrometer. (a) Schematic illustration describing how a negative arctangent spectral phase applied by the SLM, affects the group delay of the pulse. (b) Linear reflection measurement (colorful area), collected from a monolayer $\text{WSe}_\text{2}$, using the ultra-broadband pulse ($695-1030nm$). An arctangent phase mask, applied by the SLM (dashed orange), centered at the $A1s$ exciton resonant frequency $E_{X_{1s}}$ and linewidth $\gamma_{X_{1s}}$. The collected FWM signal (dark blue, $600-685nm$) scaled by a factor of $\times5\cdot10^{-6}$ to match the linear spectral amplitude. (c) Illustration of an intra-pulse FWM process in which spectral counterparts interact nonlinearly with the $A1s$ exciton. The $A1s$ state lays below the bandgap, separated by a binding energy of $E_b$ \cite{He2014}.}
    \label{fig1}
\end{figure}

\textcolor{black}{In our experiments, we use a spatial light modulator (SLM) based pulse shaping apparatus, illustrated in Fig. \ref{fig1}. The SLM, a one-dimensional liquid crystal pixelated array, is situated in the Fourier plane of a 4f system \cite{Weiner2000}, where the sub-$10fs$ ultra-broadband pulse is spatially dispersed into independent co-aligned spectral components.  The SLM allows us to temporally shape the spectral phase of the pulse $\phi_{SLM}\left(\lambda\right)$, as illustrated in Fig.\ref{fig1}.a. In addition, the Fourier plane also functions as an ultra-steep edge filter, used to sharply truncate the blue end of the spectrum.  
Following the tight focus of a mirror objective, the shaped and truncated pulse (1.2-1.77eV) interacts with a $\text{WSe}_\text{2}$ mono-layer. The linear reflection measurements we have acquired using our ultra-broadband pulse (displayed in Fig.\ref{fig1}.b), show a resonant line shape around $1.66[eV]$, confirming the photo-excitation of 
the $1s$ resonant state of the A-exciton \cite{Gu2019, Mueller2018}. 
Apart from the linear interaction, the pulse also interacts nonlinearly with the monolayer, generating various intrapulse wave-mixing processes, such as SFG and FWM (see Fig\ref{fig1}.c, more details in supplementary note S2). 
After attenuating the truncated fundamental pulse with a complementary shortpass filter, we collect, via reflection, the anti-stokes FWM signal from the illuminated sample (see Fig.\ref{fig1}.b) \cite{Dudovich2002, Suchowski2013, Kravtsov2016}.} 

\begin{figure}[h]
    \centering
    \includegraphics[scale=0.46]{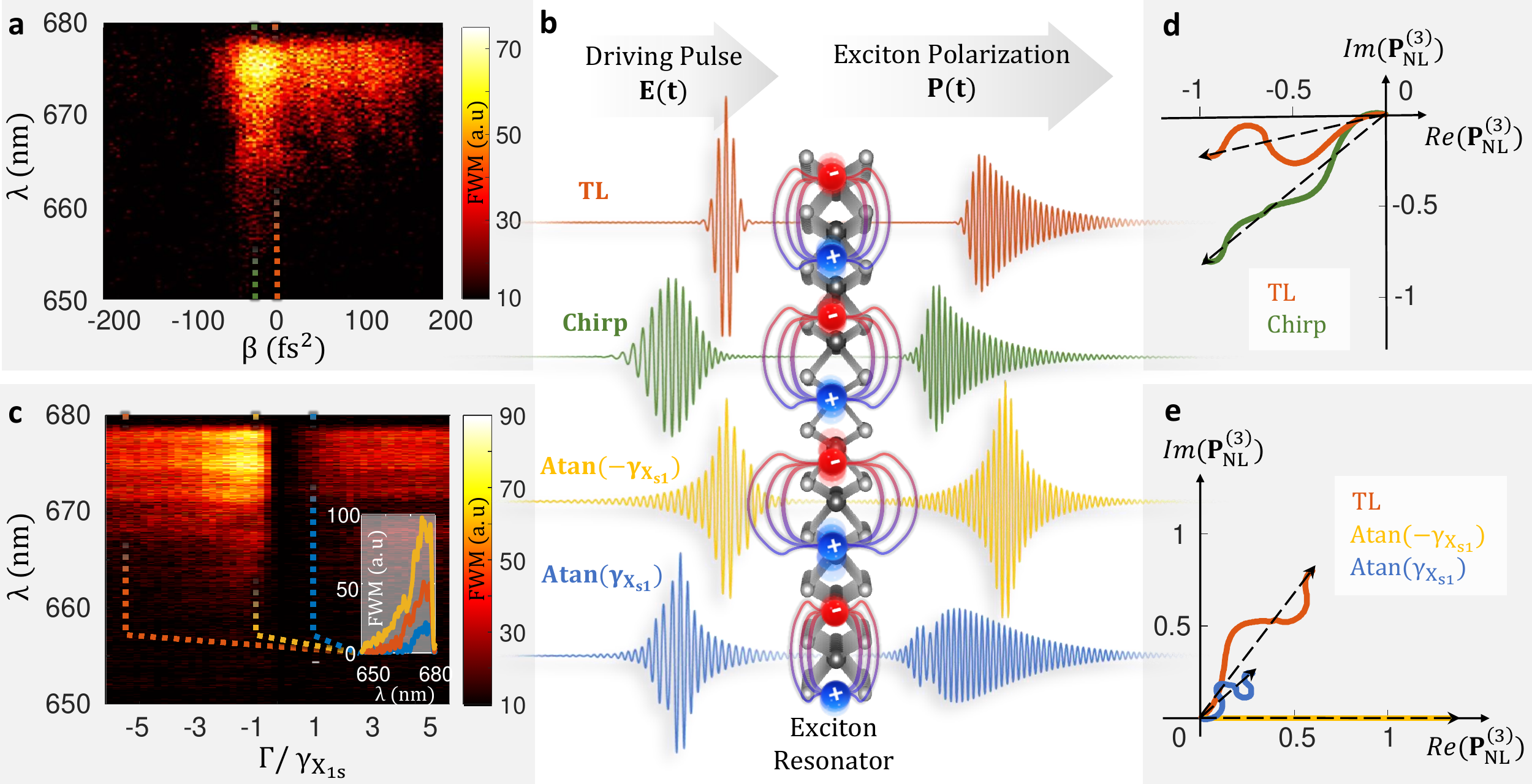}
    \caption{\textbf{The noninstantaneous resonant dynamics of the $A1s$ exciton state in monolayer $\text{WSe}_\text{2}$.} (a) The measured FWM spectrum generated by a chirped pulse, with a varying GDD of $\beta$. 
    (b) Left: An illustration of an incident time-dependent electric field, for the TL pulse (orange), a chirped pulse (green) $\beta=-20[fs^2]$, an arctangent phased pulse whose linewidth $\Gamma=-\gamma_{X_{1s}}$ (yellow) and linewidth $\Gamma=\gamma_{X_{1s}}$ (blue). 
    The incident electric field $E(t)$ interacts with the exciton oscillator, generating a polarization $P(t)$. Right: the time-dependent polarization resulting from the incoming pulse shape. The TL/chirped pulse generates a polarization with an arctangent phase (orange/green). A negative arctangent phased pulse induces a TL polarization (yellow), while a positive arctangent phased pulse induces a relatively weak polarization (blue). 
    (c) The measured FWM spectrum generated by an arctangent phased pulse centered around the $A1s$ exciton resonance, $\omega_{X_{1s}}$, while varying the spectral linewidth $\Gamma$. The plotted linewidth is normalized by the exciton's linewidth $\Gamma/\gamma_{X_{1s}}$. Inset: three selected FWM spectra for the three values of $\Gamma$ that correspond to the incoming shaped pulses from (b). (d-e) Geometrical representation of the pulse-shape dependent third-order polarization, plotted in the complex plane and reflecting the intricate multiphoton pathway interference effects. (d) Two phasors, green phasor corresponds to the chirped pulse from (b) and experimental result (dashed green line in (a)), orange phasor corresponds to TL pulse. (e) Three phasors correspond to the arctangent phases detailed in (b) and the experimental results in inset (c).}
    \label{fig2}
\end{figure}

\begin{figure}
   \centering
    \includegraphics[scale=0.45]{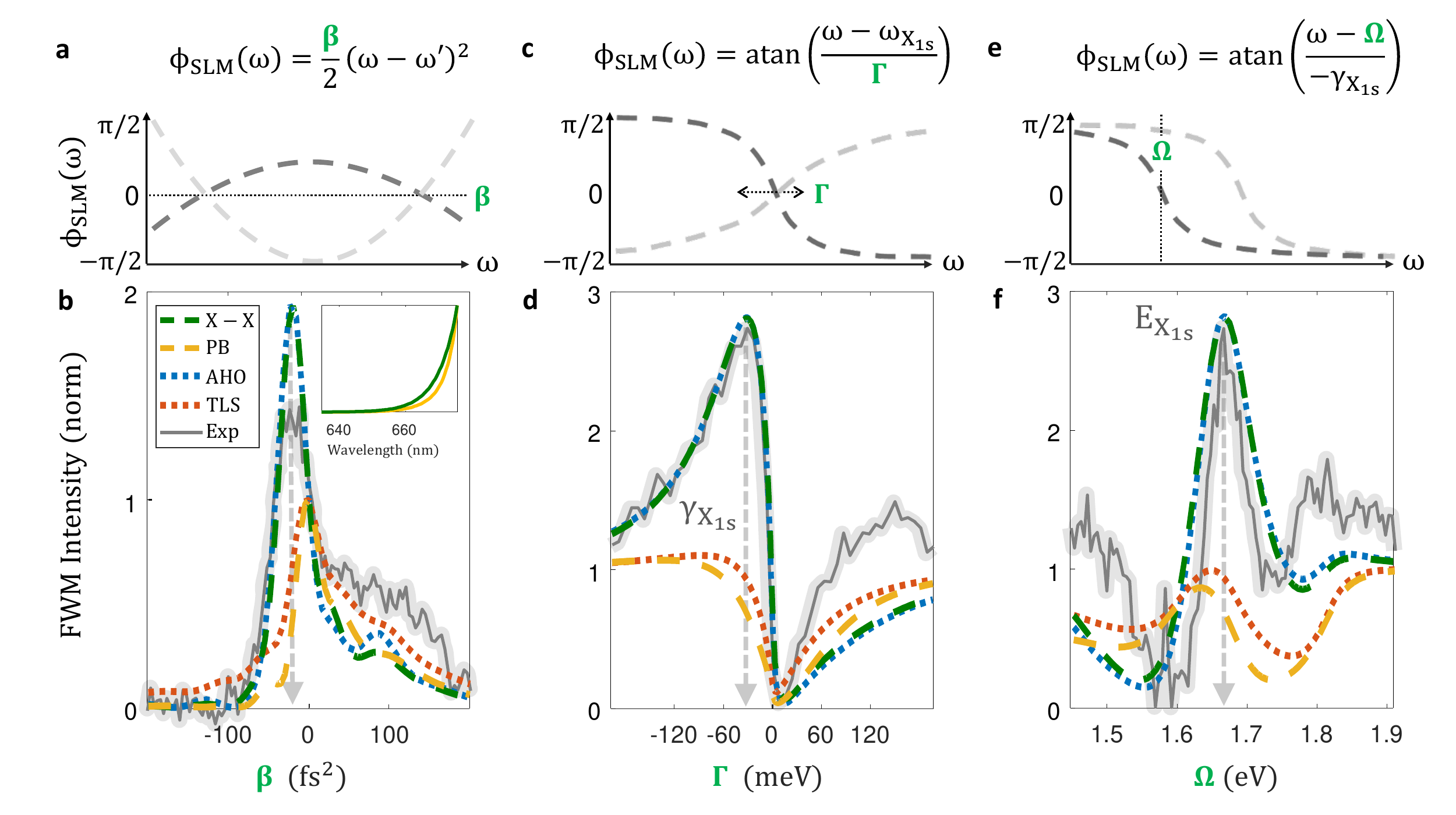}
    \caption{\textbf{FWM induced by three different phases functions, experiment Vs. three theoretical models: TMD Bloch equations of motion including Pauli blocking (PB) and exciton-exciton (X-X) interactions separately, anharmonic oscillator (AHO) and the two-level system (TLS)} (a,c,e) Analytical representation and plots of three-phase functions applied to the SLM. The varying parameter is marked in green. (b,d,f) Our experimental FWM results plotted against theoretical predictions. (a-b) FWM intensity induced by a chirped pulse, with a varying GDD of $\beta$. Inset: Theoretical prediction of FWM induced by a TL pulse, in which the two contributions (PB and X-X) give a nearly indiscernible result. (c-f) FWM intensity induced by an arctangent phased pulse. (c-d) The arctangent phase is centered around the $A1s$ exciton resonance $\omega_{X{_1s}}$, while the linewidth $\Gamma$ is varied from negative to positive values. (e-f) The arctangent phase is set with a constant linewidth $-\gamma_{X{_1s}}$, and varying central frequency $\Omega$. In both cases, (d) and (f), The FWM is maximized when $\Omega$ and $\Gamma$ reach the typical values complementary to the $X_{1s}$ inherent phase. All experimental results are scaled by a factor of 1.2 to fit the maximal enhancement of the X-X term.}
    \label{fig3}
\end{figure}

The noninstantaneous nature of resonant dynamics is characterized by a temporally asymmetric photo-induced response \cite{Dudovich2001}. Provided that the incoming pulse duration is shorter than the resonant decoherence time scale, its temporally asymmetric nature can be probed by simply varying the pulse group delay dispersion (GDD), i.e applying a parabolic spectral phase which linearly varies the instantaneous frequency of the pulse (also referred to as a linear Chirp, see Fig.\ref{fig3}.a) \cite{Chelkowski1990, Bahar2022}.
Our attempt to probe the temporal exciton dynamics of the $A1s$ transition, in ambient conditions, using SFG measurements resulted in a temporally symmetric curve. Meaning, the nonlinear response we have collected 
\textcolor{black}{
lacks the noninstantaneous resonant dynamics we would expect when interacting with the A1s exciton
, and rather resembles an instantaneous second order nonlinear response
(more details in supplementary note S2.4).}
Third-order nonlinearities, on the other hand, are independent of inversion symmetry considerations, making them suitable for measuring a wider range of ultrafast phenomena.
In the following measurements, we examined the anti-stokes FWM generated by a monolayer. 
Figure \ref{fig2}.a displays a GDD scan in which we found a temporally asymmetric trend with respect to the TL pulse, where the GDD equals zero. The FWM signal is maximal for a GDD of $\beta=-20\pm6fs^2$ and reaches a 1.4-fold enhancement with respect to the nonlinear optical response generated by a TL pulse. 
There is no doubt that the signal originates from a noninstantaneous resonant interaction, but the question remains, with which resonant level do the photons interact?

\textcolor{black}{In this experiment, we rely on the TMD Bloch equations of motion formulated by A. Knorr et al. \cite{Trovatello2020, Katsch2020, Katsch2019}. These equations serve as the foundational framework for extracting third-order parametric nonlinearity, as detailed in supplementary note S5. To simplify our analysis under weak near-infrared excitation in ambient conditions, we make several reasonable assumptions. First, we exclude contributions from exciton levels beyond $A1s$. Second, incoherent terms are disregarded since our primary focus is on observing a parametric process. Lastly, we do not account for exciton coupling to higher-order bound states, such as Biexcitons.
}

\textcolor{black}{The linear solution for the ground state to A1s dipole transition $\tilde{p}_{0}$, can be expressed as:
\textcolor{black}{\begin{equation}\begin{split}
\tilde{p}_0(\omega)= \mid \tilde{E}^*(-\omega)\mid\cdot\mid \tilde{D}^*(-\omega)\mid e^{-i(\phi_E(-\omega)+\phi_D(-\omega))},
\label{eq:Fourier_solution_linear3}
\end{split}
\end{equation} 
}
were the electric field $\tilde{E}(\omega)$ represents the shaped driving pulse at the position of the monolayer and $\tilde{D}(\omega)=-\frac{d}{\hbar}(\omega-\omega_{X} + i\gamma_X)^{-1}$ is a resonance weight function that encompasses parameters like the exciton energy $\hbar\omega_X$, dephasing rate $\gamma_X$, and optical transition matrix element $d$. According to the equations of motion, the nonlinear response under near-infrared excitation arises from two primary effects: Pauli blocking (PB) by coherent excitons and nonlinear exciton-exciton (X-X) interactions.
}

\textcolor{black}{To elucidate the third-order nonlinear contribution, we introduce a perturbation $\delta \tilde{p}$ to the linear solution and simplify the equations of motion accordingly. This perturbation analysis provides insights into the system's response beyond its linear behavior, revealing the interplay between PB and X-X interactions. The Fourier-transform of the polarization perturbation $\delta P$ takes the form:
}
\textcolor{black}{\begin{equation}\begin{split}
\delta \tilde{P}(\omega) = - \hat{d} \left(\frac{\tilde{E}^{*}(-\omega)*\tilde{p}_{0}(\omega)*\tilde{p}^{*}_{0}(-\omega)}{\hbar(\omega+i\gamma_{X} + \omega_{X})} + c.c \right) \\
- \hat{V} \left(\frac{\tilde{p}_{0}(\omega)*\tilde{p}^{*}_{0}(-\omega)*\tilde{p}_{0}(\omega)}{\hbar(\omega+i\gamma_{X} + \omega_{X})}  + c.c \right).
\label{eq:Fourier_solution_nonlinear3}
\end{split}
\end{equation} 
}
\textcolor{black}{The first contribution characterizes PB, while the second characterizes nonlinear X-X interactions.} 
Unlike narrow-band sources where the convolution simplifies to a sum over discrete mixing frequencies, our interest lies in preserving all intra-pulse four-wave combinations within our ultra-broadband source.
\textcolor{black}{It is evident that the two contributions to the nonlinearity respond differently to electric fields with varying phases, allowing us to distinguish and determine their dominance in creating third-order nonlinearity. To achieve this, we will focus on the X-X interaction and seek the pulse shape that maximizes its contribution.
}

From \textcolor{black}{Eq. (\ref{eq:Fourier_solution_nonlinear3})} it is clear that the nonlinear response arising from \textcolor{black}{$\lvert\delta\tilde{P}(\omega)\rvert$}, is maximized if \textcolor{black}{$\tilde{p}_{0}(\omega)=\lvert \tilde{p}_{0}(\omega) \rvert$.} This can be achieved by setting the electric field phase to complement the phase of the resonance weight function $\tilde{D}(\omega)$ from Eq. (\ref{eq:Fourier_solution_linear3}): \textcolor{black}{$\phi_E(\omega) = -\phi_D(\omega) = -tan^{-1}\left(\frac{\omega-\omega_X}{\gamma_{X}}\right)$.}
The arctangent phase maximizes the \textcolor{black}{dipole transition $p_0$, the linear polarization term $ \tilde{P}_0(\omega)=\tilde{p}_{0}(\omega)+\tilde{p}^{*}_{0}(-\omega)$ and most important the third-order nonlinear response generated by the X-X term In Eq.  (\ref{eq:Fourier_solution_nonlinear3}).} In Fig.\ref{fig2}.b we illustrate the relation between the incoming pulse phase and its consequential influence on the linear polarization in the time domain. Despite a small decrease in incoming pulse peak intensity caused by the negative arctangent phase, the instantaneous polarization actually increases. 

To verify these predictions  
we measured a set of arctangent phased pulses. The phase function 
$\phi_E(\omega)= tan^{-1}\left(\frac{\omega-\omega_{X}}{\Gamma}\right)$ was inserted with a varying linewidth $\Gamma$ while holding the central frequency $\omega_{{X}}$ equal to the $1s$ resonance (obtained by linear reflection as seen in Fig.\ref{fig1}.b, and verified by photoluminescence, supplementary note S1). The FWM results displayed in Fig.\ref{fig2}.c show a nearly anti-symmetric intensity trend with respect to the sign of $\Gamma$, while the maximal value is obtained for 
$\Gamma=-32\pm3 \left[meV\right]$. This value, which dictates the width of the arctangent curve, applied to the incoming electric fields' phase, matches the negative value of the linearly measured resonance linewidth $\gamma_{X_{1s}}$ from Fig.\ref{fig1}.b, and results with an enhancement factor of $2.3$ with respect to the extreme measured values of $\Gamma$, where the phase approaches a TL pulse. As $\Gamma$ increases to positive values, the FWM intensity falls dramatically until it almost vanishes. 
\textcolor{black}{The resulting multiphoton pathway interference effects observed along the GDD scan (Fig. \ref{fig2}.a) and the arctangent phase $\Gamma$-scan (Fig. \ref{fig2}.c), can be effectively visualized by employing a geometrical representation of the exciton polarization phasor in the complex plane, as illustrated in Fig. \ref{fig2}.d-e, respectively (additional examples are presented in supplementary note S6).}
A similar set of arctangent phased pulses was measured for a varying resonant frequency $\Omega$ while holding a constant linewidth of $-\gamma_{X_{1s}}$. A schematic description of the phase functions and plot of the summed FWM intensity for all three measurement sets is displayed in Fig.\ref{fig3}. The line-shape obtained by the resonance frequency scan $\Omega$ in Fig.\ref{fig3}.f reaches the maximal FWM for $\Omega=\omega_{X_{1s}}=1.66\pm4.6\times10^{-3} \left[eV\right]$. 
\textcolor{black}{These results demonstrate unambiguously that arctangent-phased pulses enable the extraction of resonant characteristics from the material with which the pulse interacts. In addition, }
they offer a unique capability to enhance or diminish third-order nonlinear polarization 
\textcolor{black}{while maintaining a high peak power in the pulse (more details in supplementary note S6.1).}
Here we note again that SFG results for similar arctangent phase scans display a nonresonant response (See supplementary note S2).


\textcolor{black}{As mentioned before, our theoretical approach simplifies the intricate dynamics of multi-resonant excitons to a basic two-level system, incorporating a first-order correction to account for X-X interaction. This simplified model effectively captures the essential aspects of ultrafast exciton dynamics when stimulated by a weak near-infrared pulse. }
In Fig.\ref{fig3}, we present our experimental FWM measurements, comparing the measurements of three varying phase functions 
\textcolor{black}{with the two nonlinear contributions predicted by Eq. (\ref{eq:Fourier_solution_nonlinear3}): the PB and X-X interactions considered separately. Notably, the X-X term aligns well with the observed trend in FWM measurements, in stark contrast to the PB term. 
\textcolor{black}{This allows us to establish a conservative estimate for the lower bound of the ratio between the X-X and PB contributions to the total FWM, determining it to be at least $2\times10^{14}$ (see supplementary note S7).}
Moreover, given that the nonlinearity stemming from the PB term can be effectively characterized using a perturbed two-level system model, and the X-X interactions exhibit dynamics akin to those observed in the classical Anharmonic Oscillator (AHO) Model, it is evident that the AHO model adequately predicts these ultrafast exciton dynamics as well.}

\begin{figure}
    \centering
    \includegraphics[scale=0.45]{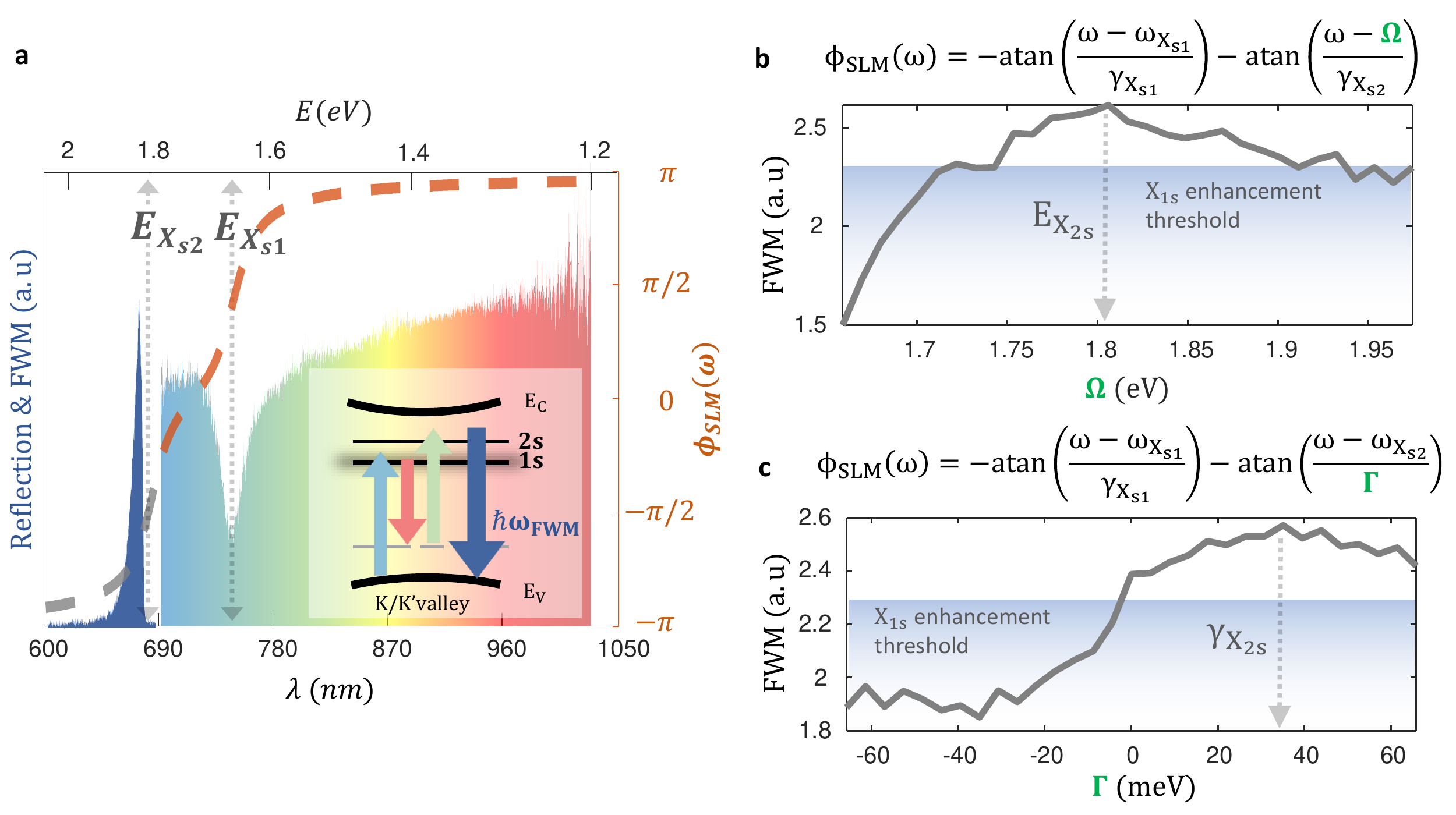}
    \caption{\textbf{FWM induced by a superposition of arctangent phases, indicating a detuned interaction with the $X_{2s}$ state}. (a) Linear reflection spectrum (colorful area) collected from a monolayer $\text{WSe}_\text{2}$, using the ultra-broadband laser. A superposition of arctangent phases applied by the SLM (dashed orange) and its analytic continuation (dashed gray), centered at the $X_{1s}$ and $X_{2s}$ resonant frequencies.  
    The collected FWM signal (dark blue). (b-c) Our experimental FWM results as a function of the superimposed arctangent central frequency $\Omega$ (b) or linewidth  $\Gamma$ (c), applied by the SLM. In both cases, the FWM is maximized for values that match the $2s$ resonant state $\Omega=\omega_{X_{s2}}$ and $\Gamma=\gamma _{X_{s2}}$.        
    }
    \label{fig4}
\end{figure}

\textcolor{black}{Having established the effectiveness of a simplified model encompassing a single resonance in capturing exciton dynamics}, we observed a slight amplification along the $\Omega$ scan (Fig. \ref{fig3}.f) for frequencies beyond our pulse bandwidth. This observation suggests potential interaction with an additional resonant state. To investigate this, we conducted a phase scan using a superposition of two arctangent functions, with one arctangent complementing the $A1s$ state while the second arctangent varied in frequency or linewidth. Remarkably, these additional scans revealed a more pronounced enhancement for frequencies corresponding to the next resonant level in the exciton Rydberg series, namely the $A2s$ state.
Despite the absence of spectral overlap between the pulse and the $A2s$ resonant frequency or its surrounding linewidth, we found that the nonlinear interaction exhibited a high sensitivity to its inherent phase. Figure \ref{fig4} illustrates two phase scans for frequency $\Omega$ and linewidth $\Gamma$, affirming that the superposition of arctangent phases allows us to control the FWM arising from multiple resonant states. Furthermore, as the tail of the inherent oscillator phase extends into our pulse bandwidth, our method proves highly sensitive to such detuned interactions.
We achieved a total \textcolor{black}{pulse-shape-based} enhancement factor of $2.6$ with respect to the TL phase. The values yielding optimal enhancement, resonant frequency $\omega_{X_{2s}}=1.81\pm0.01 [eV]$ and linewidth $\gamma_{X_{2s}}=35\pm4 [meV]$, qualitatively align with literature values for typical 2s state parameters under ambient conditions \cite{Brem2019}.

\section{Conclusion}\label{sec13}

In conclusion, \textcolor{black}{our single pulse FWM apparatus, coupled with the sub-10fs pulse-shaping capabilities,} has demonstrated the dominant role of the exciton resonance in the third-order nonlinear susceptibility $\chi^{(3)}$. This dynamic control over exciton coherence allowed us to steer the resulting multiphoton pathways interference from near destruction to a nontrivial enhancement, surpassing the maximal nonlinear yield achievable with a TL pulse. 
\textcolor{black}{Furthermore,} our measurements provided good estimations for the resonant frequency and linewidth of the band-edge A exciton 1s-orbital state, as well as valuable insights into the less accessible 2s state under ambient conditions. \textcolor{black}{The application of} the TMD Bloch equations of motion, incorporating contributions from PB and especially X-X interactions, was crucial in predicting the key features of ultrafast coherent exciton \textcolor{black}{polarization} dynamics. \textcolor{black}{Our results confirm that\textcolor{black}{, in moderate exciton densities ($\approx10^{12} cm^{-2}$) excited under ambient conditions,} the X-X interaction plays a significantly larger role in the FWM signal generation than the PB term, with the X-X interaction contributing by more than 14 orders of magnitude compared to PB.} Our ability to distinguish between these contributions to the overall nonlinear polarization, using only the spectral phase degree of freedom \textcolor{black}{of a single pulse}, underscores the precision and control enabled by pulse shaping.


\textcolor{black}{Our study uniquely captures ultrafast exciton dynamics within the regime where incoherent phonon interactions are not yet dominant, specifically using sub-10 fs pulses. The extremely short light-matter interaction timeframe, shorter than a single optical phonon cycle, may limit the influence of incoherent exciton-phonon scattering,  potentially making the FWM response primarily sensitive to coherent processes. While incoherent processes, such as phonon-assisted intervalley scattering into spin- and momentum-forbidden dark states \cite{madeo2020}, are not expected to significantly affect our results, we acknowledge that other resonant states near the A1s exciton, such as trions \cite{Riche2020} or higher-lying excitonic states \cite{Yao2020, Lin2021, Hao2017, Katsch2019}, could play a role. Additionally, we recognize the potential relevance of coherent anti-Stokes Raman scattering (CARS) \cite{Oron2002b, Wang2022} under certain pulse shapes, although detailed calculations of such effects remain outside the scope of this work.
}

Our initial demonstration of selective control between two resonant states through spectral phase manipulation highlights the potential to govern transient exciton populations within specific excited states. This approach, which we plan to substantiate further through experimental confirmation, represents a significant advancement in the field. 
\textcolor{black}{Looking ahead, we advocate for continued exploration of pulse shaping in the study of TMDs, particularly by integrating this method with other traditional methods such as pump-probe \cite{Dudovich2002a} or multidimensional spectroscopy. Such experimental capabilities could enhance the precision and depth of the study of coherent as well as the subsequent incoherent exciton polarization dynamics.}

\backmatter

\section{Methods}\label{sec11}
\subsection{Sample preparation}
$\text{WSe}_\text{2}$, obtained from HQ Graphene, was exfoliated onto commercially available polydimethylsiloxane (PDMS) gel-pak. Single-layer flakes of $\text{WSe}_\text{2}$ were identified using optical contrast. Subsequently, suitable flakes were deterministically stamped from PDMS to oxidized Si (Si/SiO2: 90nm) at room temperature and ambient conditions. To rule out interference effects from the thin SiO2 cavity, we verified our results with an additional sample consisting of a monolayer $\text{WSe}_\text{2}$ on Sapphire.

\subsection{Ultrabroad-band pulse shaping apparatus}
A mode-locked Ti:sapphire femtosecond laser system (Venteon; $7fs$, $650-1100nm$, $80MHz$) was used as the fundamental beam with linear polarization. The beam propagates through a 4f Pulse shaping system in which a spatial light modulator (Jenoptik; SLM-S640d) was placed in the Fourier plane, where we also placed mechanical barriers serving as edge filters. After propagating through a GVD mirror pair (Venteon; DCM11), The beam was focused by a mirror objective (Pike; NA 0.78) to pump the sample ($40\mu W$). The generated nonlinear signal was collected through the same objective via reflection. After propagating through two short-pass filters (Semrock;  FF02-694/SP-25), and filtering out the fundamental beam, the emitted nonlinear signals were focused onto a spectrometer (Princeton instruments; Pixis:400BReXcelon). The zig-zag axis orientation of the $\text{WSe}_\text{2}$ 2D crystal was found using a lambda-half polarizer placed in front of the objective. The described apparatus was assembled aiming for minimum dispersion, such that the focus plane will also serve as a TL phase plane. Fine dispersion tuning, to reach a TL pulse, was accomplished using the SLM.
\textcolor {black}{We tested all spectral phases applied during the experiment by the SLM using a nonresonant nonlinear crystal (BBO) to confirm that the observed enhancement does not result from deviations from a TL pulse.}

\bmhead{Data availability statement}
The datasets generated during and/or analyzed during the current study are available from the corresponding author on reasonable request.

\bmhead{Supplementary information}
Supplementary notes S1-S8, Figs. S1-S12 and References [1-10].

\bmhead{Acknowledgments}
\textcolor{black}{We would like to express our sincere appreciation for a fruitful discussion with Prof. Andreas Knorr, which has significantly contributed to the refinement of our work. Prof. Knorr has assistance in reassessing the adaptation of the TMD Bloch equation of motion to our experimental framework. }

\textbf{Contributions:}
U.A. and O.M. conceived and planned the experiments. U.A. E.B. and O.M. assembled the experimental system. O.M. carried out the experiments. U.A. and O.M. planned and carried out the simulations. S.D. contributed to sample preparation. O.M., M.B.S and H.S. wrote the manuscript. H.S. supervised the project. 

\section*{Declarations}
None.

\bibliography{CCFWM.bbl} 

\end{document}